\documentclass[aps,prl,showpacs,twocolumn,amsmath,amssymb]{revtex4-1}

\usepackage{graphicx}
\usepackage{dcolumn}
\usepackage{bm}

\begin{document}

\title{Continuous Coupling of Ultracold Atoms to an Ionic Plasma via Rydberg Excitation} 
\author{T.M. Weber}
\email{tweber@physik.uni-kl.de}
\author{T. Niederpr\"um}
\author{T. Manthey}
\author{P. Langer}
\author{V. Guarrera}
\author{G. Barontini}
\author{H. Ott}
\affiliation{Research Center OPTIMAS, Technische Universit\"at Kaiserslautern, 67663 Kaiserslautern, Germany}

\date{\today}

\begin{abstract}
We characterize the two-photon excitation of an ultracold gas of Rubidium atoms to Rydberg states analysing the induced atomic losses from an optical dipole trap. Extending the duration of the Rydberg excitation to several ms, the ground state atoms are continuously coupled to the formed positively charged plasma. In this regime we measure the $n$-dependence of the blockade effect and we characterise the interaction of the excited states and the ground state with the plasma. We also investigate the influence of the quasi-electrostatic trapping potential on the system, confirming the validity of the ponderomotive model for states with $20\leq n\leq 120$.
\end{abstract}


\maketitle
Several proposals have demonstrated that dressing ultracold atoms with highly excited Rydberg states can be an extremely powerful tool to tune the interactions among them \cite{1,2,3}. In particular long-range interactions would lead to novel quantum phases of matter as super-solid phases or dipolar crystals, once the dressed atoms are loaded into optical lattices \cite{4,5}. With respect to the regime of frozen Rydberg gases, where all the relevant physics can take place within a few $\mu$s \cite{6,7,8}, the onset of such exotic phases would require a much longer timescale, during which the coherent excitation must be preserved. A limitation in this case comes from the tendency of Rydberg gases to spontaneously evolve into a plasma \cite{9,10}. Hence a complete understanding of this process is fundamental in order to find the appropriate strategies to preserve the dressed states for sufficiently long times.
In this paper we report on the characterisation of the dynamics originating from the two-photon excitation of an optically trapped ultracold gas of neutral atoms to Rydberg states. We implement excitation pulses of at least a few tens of $\mu$s, i.e., sufficiently long to ensure that the Rydberg gas spontaneously evolves into a plasma. We exploit the excitation itself to continuously couple the neutral ultracold atoms to the plasma. We measure the strength and the shape of the resulting resonance lines varying the final state $nl$, with $20 \leq n\leq 120$ and with $l=0,2$. Moreover, we study the dependence of the atom-plasma dynamics on different excitation times and trapping potentials. The comparison of our results with those obtained with a rate equation model allows us to fully characterise the system and to highlight the role of the plasma-induced blockade and of the trapping potential.

Our experimental sequence starts by cooling room-temperature vapours of $^{87}$Rb atoms in a 2D-MOT. The atoms are subsequently transferred to the main chamber by a resonant push beam, loading a 3D-MOT with a rate of $2\times10^7\,$atoms/s, thus allowing to collect $10^8$ atoms after $5\,$s. A single 15\,W beam at 10.6$\,\mu$m, produced by a commercial cw CO$_2$ laser (model C55, Coherent Inc.), is focussed at the center of the 3D-MOT to a waist of 30$\,\mu$m, creating an optical dipole trap. After a dark-MOT phase of 70\,ms, $1.2 \times 10^6$ atoms at 60$\,\mu$K are left in the trap. By ramping down the power of the trapping beam to 180\,mW in about 6\,s, we drive forced evaporation, ending up with $4\times10^4$ atoms in a cigar shaped spinorial Bose-Einstein condensate in the $\left|5S_{1/2}, F=1\right\rangle$ ground state. The final trapping frequencies are $\nu=(205,205,17)\,$Hz. Unless explicitly stated we stop the evaporation at a power of 220\,mW obtaining a thermal cloud of $10^5$ atoms at 250\,nK with a density of $\cong10^{14}\,$cm$^{-3}$.
\begin{figure}[t]
\begin{center}
\includegraphics[width=0.475\textwidth]{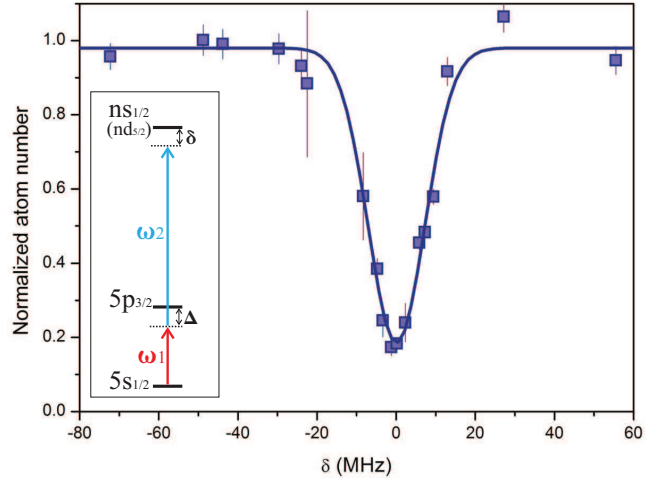}
\end{center}
\caption{(Color online) Typical resonance lineshape for a thermal cloud at 250 nK. The atoms are excited to the $27S_{1/2}$ state for 1 ms. The solid line is a Gaussian fit to the data. The inset shows the two photon Rydberg excitation scheme that we employ.} \label{fig:g1}
\end{figure} 
After the end of the evaporation we drive the transition from the $\left|5S_{1/2},F=1\right\rangle$ state to the selected $n$ Rydberg state, using the two-photon scheme depicted in the inset of Fig.$\,$1. The infra-red (IR) light at 780 nm is locked to the $\left|5S_{1/2},F=2\right\rangle \rightarrow \left|5P_{3/2},F=3\right\rangle$ transition. In this way it is always red detuned by $\Delta\simeq 2\pi \times 6.8\,$GHz from the $|5S_{1/2},F=1\rangle \rightarrow |5P_{3/2}\rangle$ transition, thus reducing the single photon scattering rate and allowing for a longer lifetime of the trapped atoms. The IR beam is collimated into the cold cloud perpendicularly to the CO$_2$ laser beam with a waist of 1\,mm and a typical power of $\simeq15\,$mW. The blue light for the second excitation step is generated by frequency doubling the light produced by a diode laser based MOPA system using a LBO crystal in a bow-tie ring cavity. The master laser can be locked in a range of about 20\,nm exploiting a combined transfer cavity and offset locking scheme, generating light from 479\,nm to 488\,nm. Once combined with the IR light, it provides the possibility to excite Rydberg states from $n=20$ to $n=150$. The actual frequency of the beam is determined with a standard EIT spectroscopy technique in a glass cell, with an accuracy of $\pm$ 5\,MHz. The blue beam is sent on the atoms collinearly with the IR one and focussed at the center of the atom cloud with a waist of $40\,\mu$m and a typical power of $\simeq120\,$mW. The applied intensities correspond to Rabi frequencies of $\Omega_1 =2\pi \times 145\,$MHz and $\Omega_2 =2\pi \times 44\,$MHz for the first and second excitation steps to the $27S_{1/2}$ state, which result in an effective overall Rabi frequency of $\Omega =\Omega_1 \Omega_2 /2 \Delta = 2\pi \times 470\,$kHz. The blue and the IR light are switched on together with variable pulse lengths, typically on the order of 1\,ms. The excitation of Rydberg atoms is revealed by the reduction of the number of trapped atoms at the end of the two-photon pulse, when scanning the frequency of the blue beam across the atomic transition. A typical line shape is shown in Fig.$\,$1.
\begin{figure}
\begin{center}
\includegraphics[width=0.475\textwidth]{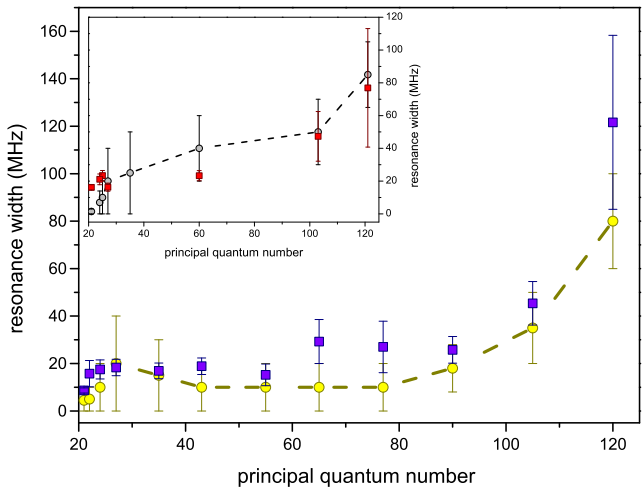}
\end{center}
\caption{(Color online) Measured resonance widths for different $ns$ states. Data (squares) are compared with the solution of the rate equation model explained in the text (circles). In the inset the same quantities are shown for $nd$ states. Note, that for $n<27$ the power of the blue excitation laser significantly drops.
} \label{fig:g3}
\end{figure}
In Fig.$\,$2 we show the measured widths of the resonances as a function of the principal quantum number $n$. 
Notably, they are up to six orders of magnitude larger than the natural linewidths ($\simeq$ 80\,kHz for $n=27$ and $\simeq$ 300\,Hz for $n=120$ \cite{11}). Moreover they show a non-trivial dependence on $n$. These features cannot be attributed to Doppler or saturation effects since, for the given experimental parameters, they are in the order of 100\,kHz. Furthermore it has been repeatedly demonstrated that the bulk excitation of Rydberg atoms in the ultracold regime rapidly leads to a series of secondary effects like fast ionization by collisions or radiation \cite{9,14}, production of plasmas \cite{21,22,23} or blockade effects \cite{24,33}. In our case, on the timescale of the excitation this secondary effects are certainly present and a complete understanding of the observed features necessarily requires to take into account all of them. For these reasons we analyse our data starting from the following rate equation model:  
\begin{eqnarray}
\dot{N}_g&=&W_{gr}\xi(N_r-N_g)+W_{ge}(N_e - N_g)\nonumber \\
&-&K_{HM} N_gN_r+K_{PI}N_eN_r+\Gamma_e N_e-\gamma_gN_g \nonumber\\ 
\dot{N}_e&=&W_{ge}(N_g - N_e)+W_{er}(N_r-N_e) \nonumber\\
&-&K_{PI}N_eN_r-\Gamma_eN_e\nonumber\\
\dot{N}_r&=&W_{gr}\xi(N_g - N_r)+W_{er}(N_e-N_r)\nonumber \\
&-&K_{BB}N_r-K_{RR}N_r-K_{HM} N_gN_r\nonumber\\
&-&K_{PI}N_eN_r-\Gamma_rN_r \nonumber\\
\dot{N}_i&=&K_{BB}N_r+\frac{1}{2}K_{RR}N_r+\frac{1}{2}K_{HM} N_gN_r \nonumber\\
&+&K_{PI}N_eN_r-\chi N_i,
\label{system}
\end{eqnarray} 
where $g$ labels the ground state atoms, $e$ the 5p state atoms, $r$ the Rydberg atoms and $i$ the ions. With  $\Gamma_e$ and $\Gamma_r$ we indicate the natural linewidths of the corresponding levels. The loss rate $\gamma_g$, which is due to the high scattering rate from the IR laser, is measured from the decay of the atom number in dependence of the excitation pulse time when the blue laser is far detuned from the transition ($\delta\simeq100\,$MHz), as shown in Fig.$\,$3a.
The coefficients $W_{ij}$ represent the excitation rates and, in our experimental regime, they are $\simeq2\pi\times$300\,Hz for the $g\rightarrow e$ transition while they range between a few Hz to a few mHz for the transitions $e\rightarrow r$, with $20<n<120$.  We calculate the two-photon coupling between the ground state and the Rydberg states using the standard formula from second-order perturbation theory \cite{26}: 
\begin{eqnarray}
W_{gr} &=& \frac{8I_1I_2}{\hbar^4 c^2 {\varepsilon_0}^2}{\left|\sum_k \frac{\widehat{d_{3k}}\widehat{d_{kg}}}{\omega_{kg}-\omega_1} + \frac{\widehat{d_{ek}}\widehat{d_{kg}}}{\omega_{ek} -\omega_2} \right|}^2\nonumber\\
&&\frac{\Gamma_r/2}{(\Gamma_r/2)^2+\Delta\omega^2},
\label{twophotoneq}
\end{eqnarray}  
where $k$ labels every intermediate state, $\widehat{d_{ji}}$ are the dipole matrix elements, $I_i$ the laser intensities and $\Delta\omega$ the overall laser linewidth. We calculate that $W_{gr}$ ranges between a few MHz ($n$=20) and a few  hundred Hz ($n$=120). All the remaining processes are related to the production of ions or to the interactions with them. When exciting Rydberg atoms ions can be produced in several ways: in our experimental circumstances the main channels are blackbody radiation (BB),  Hornbeck-Molnar ionisation (HM),  Penning ionisation (PI) and  Rydberg-Rydberg collisions (RR). At 300\,K the rate $K_{BB}$ never exceeds 500\,s$^{-1}$ for all the states that we excite \cite{16}. The HM ionisation is due to collisions between Rydberg atoms and ground state atoms that produce one Rb$_2^+$ ion and one electron. More complicated is the PI channel, in which one Rydberg atom collides with one atom in the 5p state producing one ion, one electron and one atom in the ground state. For atomic densities in the order of $10^{14}-10^{15}\,$cm$^{-3}$, the corresponding rates are $K_{HM}\simeq10^5-10^6\,$s$^{-1}$ and $K_{PI}\simeq10^6-10^7\,$s$^{-1}$ \cite{12,13}. The last ionisation channel is characterized by the well-known $n^4$ dependence of the cross-section through the rate $K_{RR}=\rho_rv(\pi a_0^2n^4)$, where $\rho_r$ is the density of the Rydberg atoms and $v$ their relative velocity \cite{17}. Every time one ion is produced, the corresponding electron leaves the trapping region extremely fast leaving an excess positive charge around the atoms. The positively charged plasma that originates from the continuous production of ions is then subjected to the so-called Coulomb explosion: the ions repel each other via the electrostatic force. The complete description of such a complicated process lies beyond the purposes of this work and we model the expansion of the plasma with an effective ion loss rate from the trapping volume $\chi=v/\sigma+\sqrt{2e^2\rho_i/(4\pi\varepsilon_0m)}/\sigma$, where $v=\sqrt{3k_BT/m}$ is the thermal velocity which we suppose to be the atomic one, $\rho_i$ the plasma density and $\sigma$ the radius of the atomic distribution. Moreover, the plasma that forms and expands produces an electric field in the region of the trapped atoms that significantly  shifts the Rydberg levels. The most striking consequence is the blockade effect, i.e., an effective reduction of the coupling between the ground state and the Rydberg state. We model this process introducing the coefficient $\xi=\Delta\omega/(\Delta\omega+B)$ in eqs.$\,$(1). Indeed, the spacially varying electric field produced by the space charge of the plasma induces a broadening of the Rydberg line that is $B=(2\sigma\rho_i/(4\pi\varepsilon_0))^2\alpha/(2\hbar)$, where $\alpha$ is the $n$-dependent atomic polarisability given by $\alpha=h\times(2.202\times10^{-7}n^6+5.53\times10^{-9}n^7)\,$Cm$^2$V$^{-1}$ \cite{32}. 
We find the linewidths calculated solving the rate equation model to be in excellent agreement with the observed ones (Fig.$\,$2). The simulated dynamics of the number of trapped atoms, Rydberg atoms and ions are shown in Fig.$\,$3b for the $27s$ state. Through the competition between the Coulomb expansion and the blockade effect the atom-plasma system rapidly evolves into a self-balanced situation where the feeding rate is continuously adjusted in order to compensate the losses of ions. In practice, for a given value of the principal quantum number $n$, the number of Rydberg atoms that are on average present in the volume is kept almost constant. Unfortunately this self-balancing effect is strongly altered by the huge losses induced by the scattering of photons from the IR laser (Fig.$\,$3a) which can not be balanced. Due to this effect the self-balanced phase has only a limited lifetime. From the solution of the rate equation model we can determine the number of Rydberg atoms $N_r$ that are present in the volume during this reduced lifetime for different $n$ states, as reported in Fig.$\,$3c. For the given trapping volume the maximum number of Rydberg atoms excited corresponds to an average distance $r=1/\sqrt[3]{\rho_r}$ that ranges from 1$\,\mu$m to 13$\,\mu$m for $27\leq n\leq 120$. In Fig.$\,$3d the $n$-dependence of the maximum number of ions present at a time is shown. 
 
\begin{figure}
\begin{center}
\includegraphics[width=0.475\textwidth]{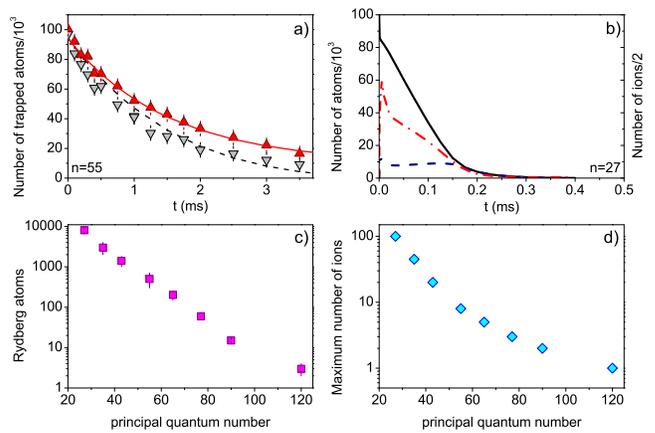}
\end{center}
\caption{(Color online) a) Typical decay measurements in (downpointing triangles) and out of resonance (uppointing triangles) together with the curves obtained as solutions of the rate equation system (2) for the $55s$ state. b) Time evolution of total number of trapped atoms (solid line), Rydberg atoms (dashed) and ions (dashed-dotted) for the $27s$ state. c) Calculated average number of Rydberg atoms present as a function of $n$. d) Calculated maximum number of ions present as a function of $n$.} 
\label{fig:g2}
\end{figure}   
       
Finally we investigate the influence of the trapping potential on the Rydberg atoms and on the plasma dynamics. The CO$_2$ laser is expected to create a repulsive ponderomotive potential for any Rydberg atom \cite{15} while for the atoms in the $5S_{1/2}$ and $5P_{3/2}$ states it creates the attractive trapping potential. We first verify the reliability of the ponderomotive assumption for the dipole potential, recording the atom losses for different trapping powers and measuring the relative shifts of the center of the resonances, as reported in Fig.4a. We observe that the AC Stark shift is effectively the same for every Rydberg state and that it is compatible with the expected theoretical value of 179\,MHz/W, as shown in Fig.4b.
\begin{figure}
\begin{center}
\includegraphics[width=0.475\textwidth]{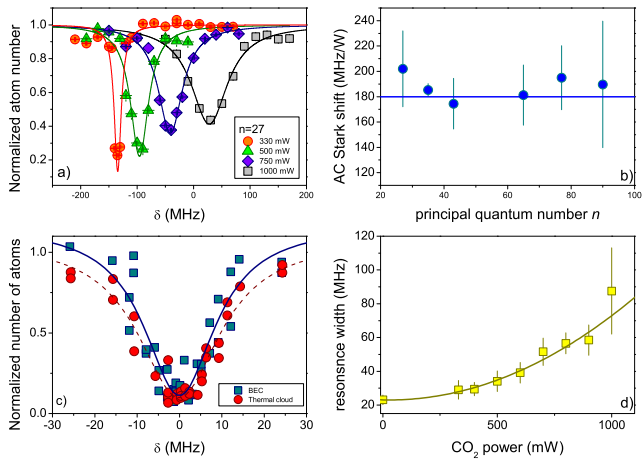}
\end{center}
\caption{(Color online) a) Typical resonance lineshapes for different powers of the CO$_2$ laser for the $27S_{1/2}$ state b) Measured AC Stark shifts for different values of $n$. The solid line is the expected state independent theoretical value of 179\,MHz/W. c) Resonance lineshapes for $n=27$ for a BEC and a thermal cloud. d) Measurement of the resonance width as a function of the CO$_2$ power for $n=43$. } \label{fig:g4}
\end{figure}
It has been demonstrated that the plasma dynamics is not directly affected by the density of the ground state atoms since the interaction is mediated mainly by the Rydberg atoms \cite{9}. We have verified this, measuring the linewidths for a BEC and a thermal cloud when the density differs by one order of magnitude while the trapping potential remains almost the same. As can be seen in Fig.4c, there is only a minimal difference between them. However, as reported in Fig.4d and as shown in Fig.4a we do observe a dependence of the linewidth on the power of the trapping laser. This effect is due to the fact that an increase of the power of the CO$_2$ laser compresses the atoms in the ground state reducing the excitation volume. In a smaller volume the ionic blockade effects are even stronger producing a more pronounced broadening and a further detriment of the transition probability.  

In summary, we have reported the two-photon excitation and subsequent fast ionization of Rydberg atoms in a unprecedented wide range of $n$ states. Elongating the excitation time to a few ms we have systematically investigated the interplay between the ultracold atoms and the formed plasma. We have carried out a detailed yet simple analysis of the system, highlighting the important role of the plasma on hampering the Rydberg excitation. We have finally characterized  the influence of a quasi-electrostatic trap on Rydberg atoms and plasma dynamics.
 
Our results have a direct impact on the schemes that aim at the dressing of ground state atoms with Rydberg states. Indeed the possible evolution of Rydberg excited samples into plasma must be taken into account, since it leads to a huge loss in the transition probability. A possible way to avoid this dynamics is the application of electric fields to remove the produced ions fastly. However, in the presence of electric fields Rydberg states are strongly mixed thus leading to population distribution over the neighbouring states \cite{31}, and hence to loss of coherence. This, together with our findings, suggests that the best way to obtain long-living plasma-free Rydberg-dressed samples would require the use of high-intensity lasers in order to reach the condition where $\delta$ is larger than the plasma-induced broadening with a decent amount of admixture. 

\begin{acknowledgments}
We acknowledge financial support by the DFG within the SFB/TRR 49 and GRK 792. V. G. and G. B. are supported by Marie Curie Intra-European Fellowships. It is a pleasure to thank P. Pillet. T. Pohl and H. Hotop for enlightening discussions. We are grateful to A. Widera for technical support.
\end{acknowledgments}


\begin{thebibliography}{999}
\bibitem{11}
Einstein-A-coefficients and dipole matrix elements are calculated by solving the time independent Schr\"odinger equation with wavefunctions derived from Numerov method and energy levels given by quantum defect theory.
\end{thebibliography}
\end{document}